\documentclass[sigconf]{acmart}
\usepackage{algorithm}
\usepackage{algorithmic}
\usepackage{pdfpages}
\usepackage{multirow}
\usepackage[export]{adjustbox} 
\usepackage{bm}
\usepackage{amsmath}
\usepackage{array}
\usepackage{booktabs}
\usepackage{footmisc}

\usepackage{amsthm}
\usepackage{graphicx}
\usepackage{enumitem}

\theoremstyle{definition}

\usepackage{subfigure}
\usepackage{url}
\usepackage{color}
\usepackage{xspace}
\usepackage{tabularx}
\usepackage{rotating}
\newcommand{\model}{FilterLLM\xspace}

\usepackage{amsmath,amsfonts,bm}

\def\eqref#1{equation~\ref{#1}}

\def\1{\bm{1}}

\def\vb{{\bm{b}}}
\def\vc{{\bm{c}}}

\def\ve{{\bm{e}}}

\def\vh{{\bm{h}}}

\def\vs{{\bm{s}}}

\def\vx{{\bm{x}}}

\def\mW{{\bm{W}}}

\DeclareMathAlphabet{\mathsfit}{\encodingdefault}{\sfdefault}{m}{sl}
\SetMathAlphabet{\mathsfit}{bold}{\encodingdefault}{\sfdefault}{bx}{n}

\def\gC{{\mathcal{C}}}

\def\gH{{\mathcal{H}}}
\def\gI{{\mathcal{I}}}

\def\gL{{\mathcal{L}}}

\def\gT{{\mathcal{T}}}
\def\gU{{\mathcal{U}}}

\DeclareMathOperator*{\argmin}{arg\,min}

\AtBeginDocument{%
  }

\setcopyright{acmlicensed}
\copyrightyear{2025}
\acmYear{2025}
\acmDOI{XXXXXXX.XXXXXXX}
\acmConference[Pre-Print]{XXX}{February, 2025}{XXX}
\acmISBN{978-1-4503-XXXX-X/18/06}

\begin{document}

\title{FilterLLM: Text-To-Distribution LLM for Billion-Scale Cold-Start Recommendation}

\author{Ruochen Liu}
\affiliation{%
  \institution{Central South University}
  \city{Changsha}
  \country{China}
}
\email{ruochen@csu.edu.cn}

\author{Hao Chen}
\affiliation{%
  \institution{City University of Macau}
  \city{Macau}
  \country{China}}
\email{sundaychenhao@gmail.com}

\author{Yuanchen Bei}
\affiliation{%
  \institution{Zhejiang University}
  \city{Hangzhou}
  \country{China}
}
\email{yuanchenbei@zju.edu.cn}

\author{Zheyu Zhou}
\affiliation{%
  \institution{Central South University}
  \city{Changsha}
  \country{China}
}
\email{8208221517@csu.edu.cn}

\author{Lijia Chen}
\affiliation{%
  \institution{Guangdong University of Finance}
  \city{Guangzhou}
  \country{China}}
\email{lijarachen@gmail.com}

\author{Qijie Shen}
\affiliation{%
  \institution{Alibaba Group}
  \city{Hangzhou}
  \country{China}}
\email{qjshenxdu@gmail.com}

\author{Feiran Huang}
\affiliation{%
  \institution{Jinan University}
  \city{Guangzhou}
  \country{China}
}
\email{huangfr@jnu.edu.cn}

\author{Fakhri Karray}
\affiliation{%
  \institution{Mohamed Bin Zayed University of Artificial Intelligence}
  \city{Abu Dhabi}
  \country{UAE}}
\email{fakhri.karray@mbzuai.ac.ae}

\author{Senzhang Wang}
\affiliation{%
  \institution{Central South University}
  \city{Changsha}
  \country{China}
}
\email{szwang@csu.edu.cn}

\renewcommand{\shortauthors}{Ruochen Liu et al.}


\begin{abstract}
Large Language Model (LLM)-based cold-start recommendation systems continue to face significant computational challenges in billion-scale scenarios, as they follow a \textbf{``Text-to-Judgment''} paradigm. This approach processes user-item content pairs as input and evaluates each pair iteratively. To maintain efficiency, existing methods rely on pre-filtering a small candidate pool of user-item pairs. However, this severely limits the inferential capabilities of LLMs by reducing their scope to only a few hundred pre-filtered candidates. To overcome this limitation, we propose a novel \textbf{``Text-to-Distribution''} paradigm, which predicts an item's interaction probability distribution for the entire user set in a single inference. Specifically, we present \textbf{FilterLLM}, a framework that extends the next-word prediction capabilities of LLMs to billion-scale filtering tasks. FilterLLM first introduces a tailored distribution prediction and cold-start framework. Next, FilterLLM incorporates an efficient user-vocabulary structure to train and store the embeddings of billion-scale users. Finally, we detail the training objectives for both distribution prediction and user-vocabulary construction. The proposed framework has been deployed on the Alibaba platform, where it has been serving cold-start recommendations for two months, processing over one billion cold items. Extensive experiments demonstrate that FilterLLM significantly outperforms state-of-the-art methods in cold-start recommendation tasks, achieving over 30 times higher efficiency. Furthermore, an online A/B test validates its effectiveness in billion-scale recommendation systems.
\end{abstract}




\maketitle

\section{Introduction}

Large language model (LLM) based cold-start recommendation remains a significant computational challenge in recommender systems. In billion-scale recommender systems, millions to billions of new products and videos are published on platforms every day~\cite{ying2018graph,chen2024macro,zhang2024multi,bei2024cpdg}, with thousands of new items being published every second. Thus, platforms must generate effective initial embeddings within seconds to ensure these items are recommended to users efficiently~\cite{wu2024survey,gao2024llm,chen2024graph,liu2023rsc}. Recently, LLMs have demonstrated significant ``Text-To-Judgement'' ability, taking user and item descriptions as input to predict whether a user will interact with an item. However, the Text-To-Judgement'' paradigm can only infer one pair at a time, resulting in linear-increased computational burden when inferring the possibilities between a user and several items, and vice versa. Consequently, designing efficient LLMs is of great priority and value for both academia and industry in cold-start recommendation.

Traditional models typically train a mapping function to generate synthetic embeddings as initial behavioral embeddings. For example, ALDI~\cite{huang2023aldi} employs knowledge distillation to minimize the differences between synthetic embeddings and behavioral embeddings. Inspired by the high performance of LLMs, a new branch of LLM-based cold-start models utilizes LLMs to simulate user-item interactions to enhance cold item embeddings. Wang et al.~\cite{wang2024large} propose using LLMs to sample both positive and negative item pairs for users from cold-start items and then training cold item embeddings by applying BPR loss to these samples. ColdLLM~\cite{huang2024large} leverages LLMs as simulators to generate multiple potential user-item interactions for each cold item, using these simulated interactions to train the cold item embeddings.

\begin{figure}[t]
\centering
\includegraphics[width=\columnwidth]{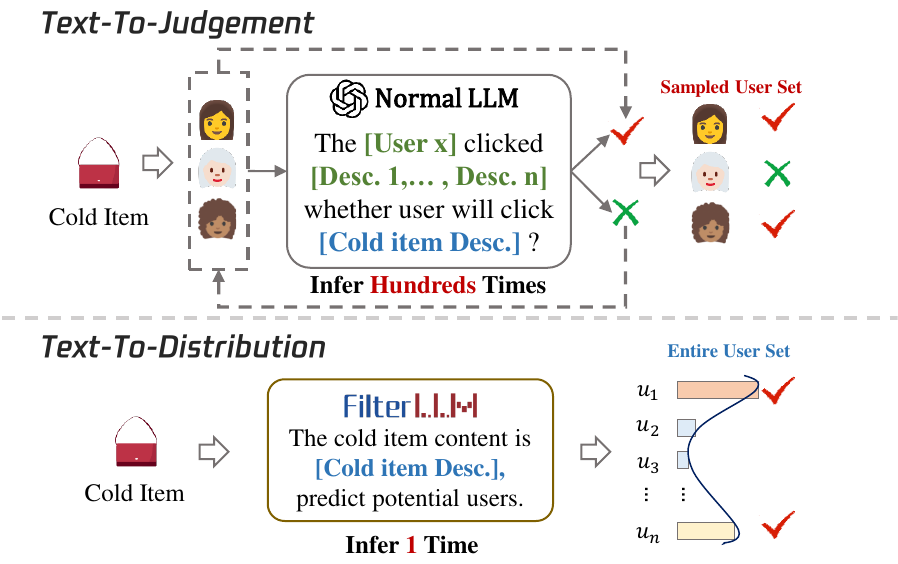}
\caption{Comparison between the ``Text-To-Judgement'' paradigm and the ``Text-To-Distribution'' paradigm.}
\label{fig:figure1}
\end{figure}

However, as shown in~\autoref{fig:figure1}, the ``Text-To-Judgement'' paradigm can only process one user-item pair at a time, requiring at least hundreds of inferences to simulate the interactions for a given item. This sequential inference paradigm brings the following limitations: \begin{enumerate}[leftmargin=*]     \item \textbf{Limited User Representation}: The ``Text-To-Judgement'' paradigm typically uses all the item content that a user has interacted with as the user context. However, this approach limits user representation, as users may have interacted with thousands of items, whereas the context window can only contain a partial subset.
\item \textbf{High Inference Time and Resource Cost}: The sequential nature of the ``Text-To-Judgement'' approach requires the LLM to check user-item pairs one by one, leading to substantial inference time and resource consumption—especially when hundreds of cold items are published every second.
\item \textbf{Small Candidate Set}: Due to sequential inference, LLM-based cold-start models must sample a small candidate set (e.g., hundreds of users) from billions of users. This limitation constrains the LLM's comprehension power, preventing it from identifying the most suitable users across the entire user set.
\end{enumerate}

As demonstrated in~\autoref{fig:figure1}, adopting the ``Text-to-Distribution'' paradigm enables the prediction of billion-scale user interaction distributions for a given item in a single inference, directly solving the linear computational burden. However, this paradigm shift introduces three critical challenges:
\begin{enumerate}[leftmargin=*]
\item \textbf{User Modeling}: Existing {description}-based user modeling approaches are constrained by the number of users they can model simultaneously. Designing a solution capable of modeling millions of users in parallel is a significant challenge.
\item \textbf{Distribution Prediction}: While LLMs excel at predicting next-word distributions, predicting distributions over billions of users presents a unique and complex challenge.
\item \textbf{LLM Training}: In natural language processing, LLMs can leverage pre-existing corpora and supervised fine-tuning (SFT) for training. Adapting this training paradigm to the recommendation domain, where the focus is on user intent distributions, remains an open problem.
\end{enumerate}

To address these challenges, we propose \textbf{\model}, a framework that extends the next-word prediction capabilities of LLMs to billion-scale filtering tasks tailored for cold-start recommendation. \model first introduces a framework to predict cold item interaction distributions from their content and refines cold item embeddings based on the predicted interaction distributions. Next, \model incorporates an efficient user-vocabulary structure to represent, train, and store billion-scale user embeddings. Finally, we present the training objectives for both distribution prediction and user-vocabulary construction, as well as the online implementation details of \model and an analysis of its computational complexity.
The contributions of our work can be summarized as follows:

\begin{itemize}[leftmargin=*]
\item We propose one single inference ``Text-to-Distribution'' paradigm for LLM-based cold-start recommendation, fundamentally resolving the iterative inference burden of the previous ``Text-To-Judgement'' approach.
\item We design and implement the \model framework, demonstrating how it effectively predicts user interaction distributions and how to train the components of \model.
\item Extensive experiments on two datasets validate the effectiveness of \model in cold-start scenarios. Online A/B test verifies that \model achieves an order-of-magnitude speed improvement over ColdLLM while enhancing cold-start performance.
\end{itemize}

\section{Related Works}
\subsection{Cold-Start Item Recommendation}
Cold-start item recommendation refers to the task of recommending newly introduced items to users despite the absence of sufficient historical interaction data for these items~\cite {zhang2025cold,liu2024fine,bai2023gorec,zhang2024logical}. 

Traditional approach to addressing the item cold-start problem relies on mapping the features of cold items into content embeddings, which are subsequently aligned with behavioral embeddings learned from warm items. This general paradigm can be referred to as ``\textit{Embedding Simulation}.'' Within this paradigm, one line of research focuses on robust co-training models which are called \textit{``dropout-based models''}, aiming to align behavioral embeddings of warm items with content-generated embeddings of cold items by employing robust training strategies~\cite{volkovs2017dropoutnet,zhu2020heater,wei2021clcrec,shi2019dropoutmethods1,xu2022dropoutmethods2}.
Another prominent direction called \textit{``generative-based model''} is knowledge alignment, where the objective is to adjust the embeddings derived from cold item content to closely align with pre-trained behavioral embeddings based on warm items~\cite{van2013deepmusic,pan2019metaemb,chen2022gar,huang2023aldi}. Then, few other efforts have paid attention to the ``\textit{Interaction Simulation}'', which generates some potential meaningful interactions between cold items and warm users/items~\cite{wang2024mutual,liu2023ucc,liu2024fine}. Representatively, USIM~\cite{liu2024fine} proposes a user behavior imagination framework with a reinforcement learning strategy for cold items,
MI-GCN~\cite{wang2024mutual} adopts pair-wise mutual information to generate in-
formative interactions for cold items before the GNN convolution.

Due to their limited natural language understanding capabilities, these methods struggle to match the recommendation performance of LLM-based approaches.

\subsection{Recommendation with LLMs}
With the rapid advancement of natural language models, LLMs have shown strong understanding capabilities~\cite{wu2024survey,zhang2025survey,hong2024knowledge,hong2024next,chen2024entity}, making them a key focus in recommendation system research~\cite{li2023large,zhao2024recommender,chang2024survey}.

Representatively, TALLRec~\cite{bao2023tallrec} applies LLMs to recommendation systems by using natural language descriptions to represent users and items. However, it faces challenges in bridging the semantic gap between natural language and user/item semantics. CLLM4Rec~\cite{zhu2024collaborative} extends LLM vocabularies with user/item ID tokens, improving semantic alignment. Additionally, methods like~\cite{rajput2023recommender, jin2023language, zheng2024adapting} use semantic IDs for item representation, particularly for sequential recommendations, reducing the need for vocabulary expansion. When focusing on the application of LLMs in cold-start scenarios, LLMs have primarily served as simulators in previous methods. Specifically, Wang et.al~\cite{wang2024large} leverages an LLM by providing it with the user’s interaction history and a pair of cold items, enabling the model to evaluate user preferences and extract semantic interactions. ColdLLM~\cite{huang2024large} employs a filtering-and-refining strategy that narrows down the user candidate set and pairs users with target cold items, using a fine-tuned LLM to query potential interactions.

However, these methods still face challenges in modeling user distribution in item cold-start scenarios and fall short of meeting the efficiency demands of real-world industrial applications.

\section{Preliminaries}
\subsection{\textbf{Notations}}From the recommendation perspective, we use $\gU$, $\gI$, and $\gH$ to represent the user set, item set, and interaction set, respectively. The item set $\gI$ can be divided into a warm item set $\gI_{w}$ (items with interaction history) and a cold item set $\gI_{c}$ (items without interaction history). For each warm item $i \in \gI_{w}$, the set of users who have interacted with item $i$ is represented by $\gU_i$. The cardinalities of these sets are denoted as $|\gU|$, $|\gI_{w}|$, $|\gI_{c}|$ and $|\gU_i|$, respectively. Additionally, let $\gC$ represent the content set of all items, where each item $i$ is associated with specific content features $\bm{c}_i$. From the large language model perspective, we use $d$ to represent the embedding dimension of the hidden state of LLM.

\subsection{{``Text-to-Judgement'' Cold-Start}}
LLM-based ``Text-to-Judgement'' cold-start methods leverage large language models to generate interactions for cold-start items, thereby addressing the item cold-start problem. The generation process can be divided into two steps.
\paragraph{Candidates Generation (billions $\rightarrow$ hundreds)}
Given the limited processing efficiency of LLMs, it is impractical for them to handle all possible user-item pairs. To address this challenge, a candidate set is first constructed to reduce the number of pairs the LLM needs to process. This process is defined as:
\begin{equation}
\mathcal{T}_i = \operatorname{Select}(\bm{c}_i, \gU, \gH),
\end{equation}
where $\mathcal{T}_i \subset \gU$  represents the candidate user set of $i$ to be processed by the LLM, and $\operatorname{Select}(\cdot)$ denotes the predefined selection rules, such as a traditional cold model~\cite{huang2024large} or random selection~\cite{wang2024large}.

\paragraph{Interaction Simulation (hundreds $\rightarrow$ tens)}
Once the candidate user set is obtained, each user is paired with the cold item $i$, and the pair is input into the LLM to predict whether it represents a true interaction. This process can be formulated as:
\begin{equation}
\hat{\gH}_i = \{(u,i)\,|\,\operatorname{LLM}(u, \bm{c}_i) =``Yes"|\, \forall u \in \mathcal{T}_i\},
\label{eq:ColdLLM}
\end{equation}
where $\hat{\gH}_i$ is the simulated interaction set of cold item $i$ and $\operatorname{LLM}(u, i) = ``Yes"$ means that there is an interaction between $i$ and $u$ simulated by large language model.

\section{Methodology}
\begin{figure*}[t]
\centering
    \includegraphics[width=\linewidth]{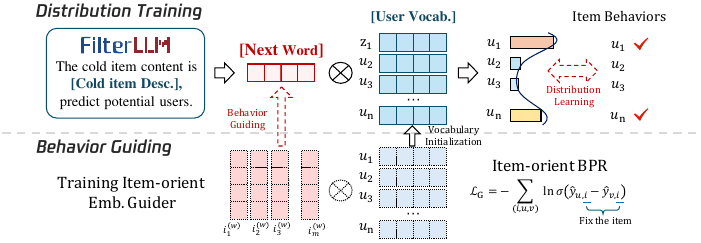}
    \caption{The overall model architecture of the \model.}
\label{fig:framework}
\end{figure*}

In this section, we first formally define the ``Text-to-Distribution'' task and illustrate how this approach tackles the cold-start problem. We then introduce a specialized LLM architecture \model to compute user distributions with the aid of a user vocabulary. After that, we elaborate on the initialization processes for the user vocabulary, along with the overall training framework for LLM-Filter. Finally, we present the online implementation details of LLM-Filter and evaluate its computational complexity compared to the state-of-the-art baseline. The overall framework is illustrated in Figure ~\ref{fig:framework}.

\subsection{Overall Framework}

Previous ``Text-to-Judgement'' methods simulate interactions by inputting a user-item pair and prompting the LLM to determine whether it represents a true interaction. However, these methods suffer from computational overhead, as they require running the LLM forward process hundreds of times to evaluate interactions for a single cold item. To address this inefficiency, we propose converting item content into a user distribution in a single LLM forward pass, and updating the cold item embeddings with this distribution. Building on this idea, we introduce our text-to-distribution task and propose the subsequent process of enhancing cold item embeddings by leveraging these distributions.

\subsubsection{\textbf{Text-to-Distribution Task}}

The goal of this task is to map the content of a cold item, represented as text, to a user distribution using a large language model. This distribution is learned by considering all historical user behaviors along with the user, item, and context information. The process can be formulated as:
\begin{equation}
\label{simulation}
    p(u|\bm{c}_i) = \operatorname{\model}(u, \bm{c}_i, \gU, \gH, \gC),
\end{equation}
where $\operatorname{\model}(\cdot)$ denotes a function that generates the user distribution for a given item in a single pass. This contrasts with
the ``Text-to-Judgment'' methods, where the process needs to be
repeated $|\gT_i|$ times, with $|\gT_i|$ being the size of the candidate set.

\subsubsection{\textbf{Enhancing Cold Embedding with Distribution}}  

Using the generated user distribution, we first sample an interaction set $\hat{\gH}_i$ for each cold item to update its embeddings, which can be formulated as:
\begin{equation}
\label{interaction-sample}
\hat{\gH}_i = \left\{ (u, i) \mid u \in \operatorname{Sample}\left( P(u \mid c_i)\right) \right\}.
\end{equation}
Here, $\operatorname{Sample}(\cdot)$represents a sampling strategy that generates a set of users \( u \) from the user distribution \( P(u \mid c_i) \) for item \( i \). Once the interactions for cold items are sampled, cold items can updated with these interactions. By leveraging the sampled interactions, the final cold embeddings for downstream tasks can be expressed as:
\begin{equation}  
\label{eq:emb-opt}
    \bm{E}_i^{(c)} = \operatorname{Opt}\left( \left[ \hat{\mathcal{H}}, \gH \right], \gI, \gU\right), \hat{\mathcal{H}}=\bigcup\limits_{i \in I_c}{\hat{\gH}_i},
\end{equation}
where \( \bm{E}_i^{(c)} \) represents the updated cold item embedding matrix, which is generated by the optimization function \( \operatorname{Opt}(\cdot) \). $\left[  \hat{\mathcal{H}} , \gH \right]$ is an interaction set which contains the original interaction $\gH$ and all sampled interaction $\hat{\mathcal{H}}$.

\subsection{\model Structure}
Building on the idea outlined above, the central challenge is how to effectively accomplish the ``text-to-distribution'' task. Traditional description-based user representations in previous methods fall short, as they are constrained by the limited number of users that can be modeled simultaneously. To overcome this limitation, we extend the original vocabulary of the LLM by incorporating a user-specific vocabulary. This extension allows the LLM to model billions of users as a distribution in a single forward pass, leveraging pre-stored user representations within the expanded user vocabulary.

\subsubsection{\textbf{User Vocabulary Expansion}}  
To support user distribution prediction, we incorporate user features into the LLM by introducing a dedicated user vocabulary, where each user ID is represented as a unique token within the model. The user vocabulary $\mathcal{V}_\text{user}$ is defined as:
\begin{equation}
    \mathcal{V}_\text{user} = \{ u \mid u \in \gU \}, \quad \bm{z}_u = \bm{O}_u \bm{Z}, \; \forall u \in \mathcal{V}_\text{user}.
\end{equation}  
Here, $\bm{Z} \in \mathbb{R}^{|\gU|\times d} $ represents the learned user token embeddings within the LLM, and $\bm{z}_u \in \mathbb{R}^d$ denotes the embedding of user $u$, where $\bm{O}_u \in \mathbb{R}^{|\gU|}$ is the one-hot vector corresponding to user $u$. These user tokens are exclusively utilized in the LLM's output space, enabling the model to predict user distributions without the need for explicit user descriptions in the input.

\subsubsection{\textbf{Content Prompt Encoder}}  
With the integration of extended user vocabulary, we only need to input prompts related to the item content, which greatly reduces the input length of tokens. Specifically, the input prompt is structured as follows:
\begin{quote}
\itshape
``Assuming you are a recommendation expert. An item has the following content {\color{blue}[Text]$(\vc_i)$}, please predict potential users for this item.''
\end{quote}
Then, this prompt will be mapped into the hidden space through a series of stacked self-attention and feed-forward modules, as expressed by the following equations:
\begin{equation}
\begin{aligned}
    \vh^{(l)}_{i,1:k} &= \sigma\left(\operatorname{SA}(\vh^{(l-1)}_{i,1:k}) \mW_1 + \vb_1\right) \mW_2 + \vb_2, \\
    \vh^{(0)}_{i,1:k} &= {\operatorname{Embed}}({\vx_{i,1:k}}).
\end{aligned}
\end{equation}
Here, $\operatorname{SA}(\cdot), \operatorname{Embed}(\cdot)$ are self-attention and embed functions, respectively. 
Additionally, ${\vx_{i,1:k}} = \text{Prompt}(\bm{c}_i)$ denotes the prompt token sequence derived from the item content $\bm{c}_i$. $\vh^{(l)}_{i,1:k} \in \mathbb{R}^{k \times d}$ refers to the hidden vectors outputted by the $l$-th layer of LLM and $k$ is the length of ${\vx_{i,1:k}}$. We use the final hidden state vector $\vh^{(N)}_{i,k} \in \mathbb{R}^{ d}$ from the last layer to represent the item content prompt.

\subsubsection{\textbf{Distribution Prediction}}
After encoding the item prompt, we further transform it into a user probability distribution with the support of user vocabulary. Specifically, we add a probability distribution prediction head to the output of LLM, \( f_l : \mathbb{R}^{d} \to \mathbb{P}(J) \), which maps the item content presentation $\bm{h}_{i,k}^{(N)}$ into the probability space $\mathbb{P}(J)$. The weights of \( f_l \) are closely related to the embedding of the extended user vocabulary, which can be written as follows:
\begin{equation}
\boldsymbol{p}(u|\bm{c}_i) = f_l(u, h_{i,k}^{(N)}) = \frac{\exp{\left( \bm{h}_{i,k}^{(N)}\bm{z}_{u}^\top\right)}}{\displaystyle \sum_{v \in \gU} \exp{\left( \bm{h}_{i,k}^{(N)}\bm{z}_{v}^\top\right)}}.
\label{distribution}
\end{equation}
Then the user distribution can be formulated as $u\sim p(u|\bm{c}_i)$, which generates user $u$ based on the item content $\bm{c}_i$.

\subsection{\model Training}
In this subsection, we introduce the \model training strategy, including the user vocabulary initialization and overall training framework of \model.

\subsubsection{\textbf{Collaborative-Driven User Vocabulary Initialization}}
Introducing a user vocabulary into the LLM can significantly enhance inference efficiency. However, the challenge lies in effectively fine-tuning the billions of newly added, randomly initialized user token embeddings. To address this, we propose a collaborative-driven method for initializing user token embeddings.

Since user token embeddings are used to model the interaction between users and items, an initialization method that effectively captures these relationships is crucial. To achieve this, we leverage collaborative filtering techniques and BPR loss, which have been shown to capture such relationships~\cite{su2009survey, schafer2007collaborative, herlocker2000explaining}, to initialize the user vocabulary. However, unlike traditional user-oriented recommendation systems, which predict items for users, our approach aims to predict the user distribution for each item. This shift in focus demands that we design the loss function from the item perspective, rather than following the user-centric approach of traditional BPR. To address this, we introduce an item-oriented BPR loss for initializing the user token embeddings, as the following formulations:
\begin{equation}
\label{initialzation}
    \begin{split}
        \mathcal{L}_{\text{BPR}} &= -\sum_{i=0}^{|\gI_{w}|}\sum_{u\,\in \gU_{i}} \sum_{v\,\notin \gU_{i}} \log \sigma(\hat{y}_{{u}i} - \hat{y}_{{v}i}), \\ \hat{y}_{ui} &= \operatorname{CF}(E_u, E_i, \gH).
    \end{split}
\end{equation}
Here, $\sigma$ denotes the sigmoid function, and $\operatorname{CF}(\cdot)$ refers to a collaborative filtering method~\cite{he2020lightgcn, wang2019ngcf, rendle2009bprmf}. In this work, we employ a graph-based method~\cite{he2020lightgcn} to capture high-order relationships between users and items. $\bm{E}_u \in \mathbb{R}^{|\gU|\times d}$ and $\bm{E}_i \in \mathbb{R}^{|\gI_{w}|\times d}$ are behavior embedding matrices for users and items, respectively, which are learned during training. $\bm{E}_i$ will serve as the item embedding guider in the subsequent \model training.

After obtaining the embedding matrices, the user token embeddings can be initialized as follows:
\begin{equation}
\bm{z}_{u} \leftarrow \bm{e}_{u}, \quad \bm{e}_{u} = \bm{O}_u \bm{E}_u,
\end{equation} 
where $\bm{O}_u \in \mathbb{R}^{|\gU|}$ is the one-hot vector corresponding to user $u$.

\subsubsection{\textbf{Distribution Learning}}
To optimize the user probability distribution, we aim to ensure that the probability of a positive user who has interacted with the item is as high as possible compared to a negative user who has not. Additionally, we strive to align the probability calculation method with next-token prediction to maintain consistency with the LLM. Therefore, we adopt the log-softmax loss~\cite{covington2016deep,wu2024bsl} to optimize the user prediction distribution. Specifically, the loss can be expressed as:
\begin{equation}
\begin{aligned}
&\mathcal{L}_{\text{Distrib}} = \\
& -{\displaystyle\sum_{i \in \gI_{w}}} \frac{1}{|\gU_{i}|} {\displaystyle\sum_{u \in \gU_{i}}} \bigg[ 
    \log \frac{\exp{\left( \bm{h}^{(N)}_{i,k}\bm{z}_u^{\top}\right)}} 
    {\displaystyle\sum_{u \in \gU_{i}}\exp{\left(  {\bm{h}^{(N)}_{i,k}}\bm{z}_u^{\top} \right)} + {\displaystyle\sum_{v \in \gU_{i}^{-}}}
    \left[\exp{\left(\bm{h}^{(N)}_{i,k}\bm{z}_v^{\top} \right)}\right]}
\bigg],
\label{log-softmax}
\end{aligned}
\end{equation}
where \( \gU_{i}^{-} \) represents the negative user set for item \( i \). When the user scale is excessively large, \( \gU_{i}^{-} \) can be constructed through sampling, transforming Eq. (\ref{log-softmax}) into a sampled log-softmax loss~\cite{blanc2018adaptive, yao2021self, wu2024effectiveness} which effectively reduces computational overhead.

Meanwhile, we observe that Eq. (\ref{log-softmax}) can be rewritten in the following vectorized form:
\begin{equation}
\begin{split}
    \mathcal{L}_{\text{Distrib}}= 
    -\sum_{i \in \mathcal{I}_w}\frac{1}{|\gU_i|}\log\left(\frac{\exp{\left( \bm{h}_{i,k}^{(N)}\bm{Z}^\top\right)}}{ \exp{\left( \bm{h}_{i,k}^{(N)}\bm{Z}^\top\right)}\boldsymbol{e}^\top}\right)\bm{y}^\top.
    \label{vec-softmax}
\end{split}
\end{equation}
Here, \( \bm{y} \in \mathbb{R}^{|\gU|} \) is a multi-hot encoding, where each element \( y_u \) is 1 if user \( u \) has interacted with item \( i \) and $\boldsymbol{e}\in \mathbb{R}^{|\gU|}$ is all-ones vector. As shown in Eq. (\ref{vec-softmax}), this approach enables the LLM to capture all historical interactions of an item in a single forward pass of LLM. As a result, only \( |\gI_w| \) forward passes are needed per training epoch, significantly improving training efficiency.

\subsubsection{\textbf{Behavior Guiding}}
In the early stages of training, LLMs, primarily designed for natural language processing tasks, often struggle to effectively interpret the user vocabulary initialized with collaborative information. To address this, we guide the LLM to imitate the behavior patterns observed in user-item interactions from the embedding perspective, enabling the capture of higher-order user-item relationships that LLMs typically miss. This approach helps the model overcome the initial misalignment between its natural language focus and collaborative data.
Specifically, the behavior guiding loss can be calculated as follows:
\begin{equation}
    \mathcal{L}_{\text{Guid}} = \sum_{i \in \gI_{w}} \left\| \bm{h}_{i,k}^{(N)} - \bm{O}_i \bm{E}_i \right\|_2^2,
\end{equation}
where $\bm{E}_i$ is the item matrix trained by Eq. (\ref{initialzation}) alongside with $\bm{E}_u$ and $\bm{O}_i\in \mathbb{R}^{|\gI_w|}$ is the one-hot vector corresponding to item $i$.

\subsubsection{\textbf{Efficient Tuning}}
The whole loss function can be calculated as follows:
\begin{equation}
    \gL = \gL_\text{Distrib} +\lambda \cdot \gL_\text{Guid},
\end{equation}
where $\lambda$ is the hyperparameter that controls the strength of behavior guiding, during model training, the original weights of the large language model are kept fixed, and only the newly introduced parameters, including LoRA~\cite{hu2021lora} layers and our user vocabulary embeddings, are fine-tuned. The whole training process can be seen in Appendix~\ref{Pseudocode}.

\subsubsection{\textbf{Inference}}
For convenience, we use the greedy sampling strategy during inference. Specifically, we select K users most likely to interact with the cold item as the sampled interactions. Then the Eq. (\ref{interaction-sample}) can be formulated as:
\begin{equation}
\hat{\gH}_i = \left\{ (u, i) \mid u \in \operatorname{Top}_K\left(p(u\mid c_i)\mid\forall u\in\mathcal{U}\}\right)\right\},
\end{equation}
where $\operatorname{Top}_K$ denotes a top-$K$ selection function. Following this, the cold item embeddings are optimized using Eq. (\ref{eq:emb-opt}).
\subsection{Industrial Deployment}
\model has demonstrated its effectiveness by providing helpful cold-start recommendations on the e-commerce platform. In this subsection, we detail the industrial deployment of \model.

\paragraph{\textbf{Deployment Architecture}}
\begin{figure}[th]
    \centering
    \includegraphics[width=\linewidth, trim=0cm 0cm 0cm 0cm,clip]{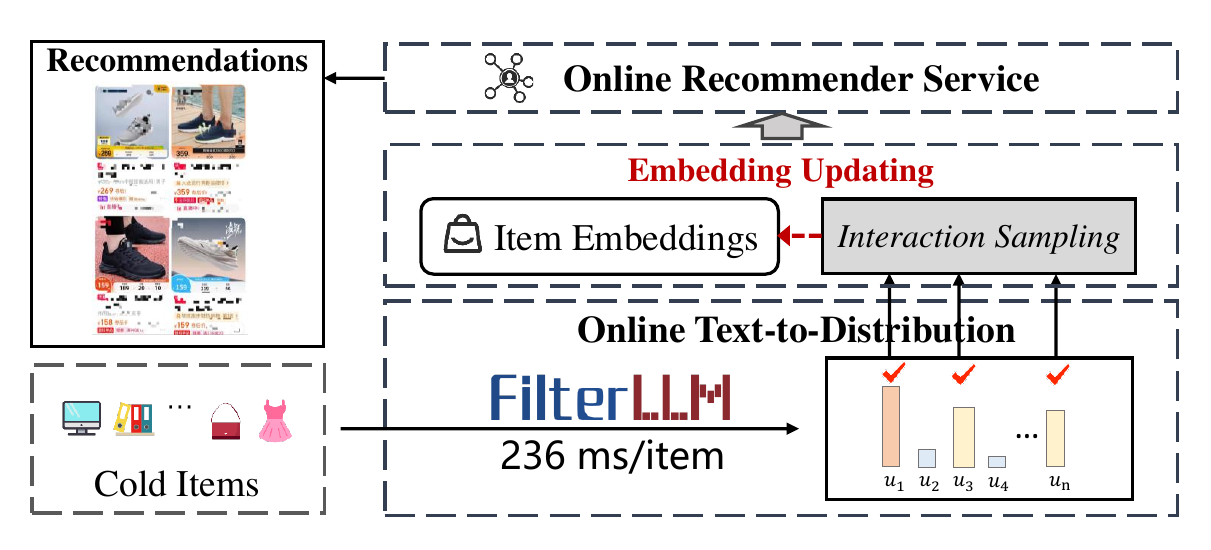}
    \caption{The overall pipeline of \model deployment.}
    \label{fig:onlineArch}
\end{figure}
As shown in Figure~\ref{fig:onlineArch}, our overall framework for \model consists of two primary components: (i) online text-to-distribution computation and (ii) online recommender serving. When new items are uploaded to the platform, \model is first used to predict the user distribution with item descriptions. 
Based on users with the highest predicted probabilities from this distribution, the most relevant user-item interactions are then sampled and fed into the embedding updating center. Since these interactions are sampled rather than actual behaviors, only the embeddings of cold items are updated. Finally, the updated embeddings are transmitted to the online recommender service.

\paragraph{\textbf{Discussion}} 
We evaluate our \model against the state-of-the-art ``Text-to-Judgement'' method ColdLLM~\cite{huang2024large} across two main aspects in industrial applications, containing Interaction Generation and Embedding Update.

\begin{enumerate}[leftmargin=*]
    \item \textbf{Interaction Generation:} The time complexity ratio between ColdLLM and \model for interaction generation is given by $\frac{L_{2}^{2}+L_{2}d}{L_{1}^{2}+L_{1}d}K$, where $L_{1}$ and $L_{2}$ denote the average token lengths of \model and ColdLLM, 
    $K$ represents the sampled user count in ColdLLM. Empirical results with LLaMA-7B show that \model achieves average {236}\, ms per cold item, yielding a {34.75}$\times$ speedup compared to ColdLLM's 8201 ms. More details about time complexity can be seen in Appendix ~\ref{Complexity Analysis}. 
    \item \textbf{Embedding Updating:} \model and ColdLLM follow the same embedding update process in industrial applications, leveraging generated interactions to optimize cold item embeddings within approximately 120 ms.
\end{enumerate}

In conclusion, \model shows notable speedup in interaction generation compared to ColdLLM, while embedding updates are similar in both models.

\section{Experiments}

\begin{table*}[htbp]
  \centering
  \caption{Results on CiteULike and ML-10M. The best and second-best results are highlighted in bold font and underlined.}
  \vspace{-0.4em}
    \begin{tabular}{cl|cc|cc|cc|cc|cc|cc}
    \toprule
    \multicolumn{2}{c|}{\multirow{3}[2]{*}{Method}} & \multicolumn{4}{c|}{Overall Recommendation} & \multicolumn{4}{c|}{Cold Recommendation} & \multicolumn{4}{c}{Warm Recommendation} \\
    \multicolumn{2}{c|}{} & \multicolumn{2}{c|}{CiteULike} & \multicolumn{2}{c|}{ML-10M} & \multicolumn{2}{c|}{CiteULike} & \multicolumn{2}{c|}{ML-10M} & \multicolumn{2}{c|}{CiteULike} & \multicolumn{2}{c}{ML-10M} \\
    \multicolumn{2}{c|}{} & \multicolumn{1}{c}{Recall} & \multicolumn{1}{c|}{NDCG} & \multicolumn{1}{c}{Recall} & \multicolumn{1}{c|}{NDCG} & \multicolumn{1}{c}{Recall} & \multicolumn{1}{c|}{NDCG} & \multicolumn{1}{c}{Recall} & \multicolumn{1}{c|}{NDCG} & \multicolumn{1}{c}{Recall} & \multicolumn{1}{c|}{NDCG} & \multicolumn{1}{c}{Recall} & \multicolumn{1}{c}{NDCG} \\
    \midrule
    \multirow{12}[6]{*}{\begin{sideways}MF\end{sideways}} & DropoutNet & 0.1032  & 0.0817  & 0.1308  & 0.1285  & 0.2608  & 0.1543  & 0.0155  & 0.0112  & 0.1821  & 0.1099  & 0.3056  & 0.1991  \\
          & MTPR  & 0.1197  & 0.0885  & 0.1122  & 0.1146  & 0.2662  & 0.1531  & 0.0320  & 0.0205  & 0.1948  & 0.1095  & 0.2621  & 0.1774  \\
          & Heater & 0.1148  & 0.0887  & 0.1214  & 0.1181  & 0.2538  & 0.1463  & 0.0709  & 0.0397  & 0.1788  & 0.1081  & 0.2827  & 0.1819  \\
          & CLCRec & 0.1260  & 0.1088  & 0.0880  & 0.0776  & 0.2740  & 0.1661  & 0.0620  & 0.0353  & 0.2350  & 0.1532  & 0.2063  & 0.1251  \\
          \cmidrule{2-14} & DeepMusic & 0.1462  & 0.1197  & \underline{0.1413}  & \underline{0.1409}  & 0.2273  & 0.1357  & 0.0608  & 0.0276  & \underline{0.2884}  & \underline{0.1953}  & 0.3295  & 0.2192  \\
          & MetaEmb & 0.1312  & 0.1184  & \underline{0.1413}  & \underline{0.1409}  & 0.2633  & 0.1569  & 0.0078  & 0.0043  & \underline{0.2884}  & \underline{0.1953}  & 0.3295  & 0.2192  \\
          & GAR   & 0.1445  & 0.1123  & 0.0193  & 0.0214  & 0.2353  & 0.1331  & 0.0303  & 0.0235  & 0.2272  & 0.1483  & 0.3201  & 0.2146  \\
          & ALDI  & 0.1665  & 0.1267  & 0.1007  & 0.0934  & 0.2391  & 0.1364  & 0.0296  & 0.0168  & \underline{0.2884}  & \underline{0.1953}  & 0.3295  & 0.2192  \\
\cmidrule{2-14}          & USIM  & 0.1724  & 0.1306  & \underline{0.1413}  & \underline{0.1409}  & 0.2459  & 0.1402  & 0.2790  & 0.0223  & \underline{0.2884}  & \underline{0.1953 } & 0.3295  & 0.2192  \\
          & Wang et.al & 0.1702  & 0.1311  & 0.1297  & 0.1211  & 0.2322  & 0.1416  & 0.0507  & 0.0297  & 0.2876  & 0.1944  & 0.3136  & 0.1980  \\
          & ColdLLM & \underline{0.1838}  & \underline{0.1406}  & 0.1289  & 0.1268  & \underline{0.3119}  & \underline{0.1850}  & \textbf{0.1195} & \textbf{0.0991} & 0.2247  & 0.1431  & \underline{0.3429}  & \underline{0.2367}  \\
\cmidrule{2-14}          & \model & \textbf{0.2128} & \textbf{0.1700} & \textbf{0.1417} & \textbf{0.1435} & \textbf{0.3486} & \textbf{0.2221} & \underline{0.0990}  & \underline{0.0625}  & \textbf{0.2899} & \textbf{0.1960} & \textbf{0.3444} & \textbf{0.2425} \\
    \midrule
    \multirow{13}[6]{*}{\begin{sideways}NGCF\end{sideways}} & DropoutNet & 0.1194  & 0.0947  & 0.1316  & 0.1271  & 0.2828  & 0.1690  & 0.0485  & 0.0248  & 0.2019  & 0.1211  & 0.3103  & 0.1971  \\
          & MTPR  & 0.1133  & 0.0849  & 0.1202  & 0.1188  & 0.2693  & 0.1568  & 0.0309  & 0.0199  & 0.1848  & 0.1064  & 0.2831  & 0.1843  \\
          & Heater & 0.1529  & 0.1206  & 0.1239  & 0.1245  & 0.2894  & 0.1728  & 0.0397  & 0.0226  & 0.2218  & 0.1355  & 0.2873  & 0.1914  \\
          & CLCRec & 0.1400  & 0.1228  & 0.1022  & 0.0955  & 0.2894  & 0.1760  & 0.0423  & 0.0270  & 0.2645  & 0.1757  & 0.2401  & 0.1534  \\
          \cmidrule{2-14} & DeepMusic & 0.1661  & 0.1496  & 0.1449  & 0.1488  & 0.2494  & 0.1502  & 0.0643  & 0.0365  & \underline{0.3335}  & \underline{0.2249}  & 0.3362  & 0.2309  \\
          & MetaEmb & 0.1584  & 0.1438  & 0.1449  & 0.1488  & 0.2905  & 0.1723  & 0.0632  & 0.0263  & \underline{0.3335}  & \underline{0.2249}  & 0.3362  & 0.2309  \\
          & GAR   & 0.1620  & 0.1263  & 0.0182  & 0.0209  & 0.2738  & 0.1656  & 0.0275  & 0.0219  & 0.2848  & 0.1840  & 0.3145  & 0.2141  \\
          & ALDI  & 0.1731  & 0.1509  & 0.1450  & 0.1489  & 0.2824  & 0.1556  & 0.0488  & 0.0310  & \underline{0.3335}  & \underline{0.2249}  & 0.3362  & 0.2309  \\
\cmidrule{2-14}          & USIM  & 0.1611  & 0.1451  & 0.1449  & 0.1488  & 0.2534  & 0.1478  & 0.0261  & 0.0217  & \underline{0.3335}  & \underline{0.2249}  & 0.3362  & 0.2309  \\
          & MI-GNN & 0.1510  & 0.1322  & 0.1338  & 0.1337  & 0.0049  & 0.0021  & 0.0060  & 0.0045  & 0.3179  & 0.2147  & 0.3272  & 0.2277  \\
          & Wang et.al & 0.1658  & 0.1368  & 0.1444  & 0.1472  & 0.2221  & 0.1369  & 0.0752  & 0.0485  & 0.2887  & 0.1983  & \underline{0.3480}  & \underline{0.2327}  \\
          & ColdLLM & \underline{0.2304}  & \underline{0.1776}  & \underline{0.1455}  & \underline{0.1504}  & \underline{0.3482}  & \underline{0.2090}  & \underline{0.0985}  & \underline{0.0744}  & 0.3277  & 0.2135  & 0.3379  & 0.2317  \\
\cmidrule{2-14}          & \model & \textbf{0.2585} & \textbf{0.2127} & \textbf{0.1504} & \textbf{0.1566} & \textbf{0.3880} & \textbf{0.2522} & \textbf{0.1787} & \textbf{0.1266} & \textbf{0.3359} & \textbf{0.2299} & \textbf{0.3501} & \textbf{0.2417} \\
    \midrule
    \multirow{13}[6]{*}{\begin{sideways}LightGCN\end{sideways}} & DropoutNet & 0.1027  & 0.0801  & 0.1319  & 0.1318  & 0.2580  & 0.1487  & 0.0196  & 0.0129  & 0.1808  & 0.1078  & 0.3088  & 0.2039  \\
          & MTPR  & 0.0977  & 0.0734  & 0.1122  & 0.1096  & 0.2418  & 0.1401  & 0.0283  & 0.0184  & 0.1621  & 0.0923  & 0.2674  & 0.1716  \\
          & Heater & 0.1281  & 0.0989  & 0.1236  & 0.1249  & 0.2746  & 0.1609  & 0.0668  & 0.0378  & 0.1966  & 0.1181  & 0.2880  & 0.1922  \\
          & CLCRec & 0.1215  & 0.1075  & 0.0993  & 0.0929  & 0.2569  & 0.1547  & 0.0401  & 0.0254  & 0.2325  & 0.1537  & 0.2329  & 0.1475  \\
          \cmidrule{2-14} & DeepMusic & 0.1401  & 0.1144  & 0.1458  & 0.1501  & 0.1759  & 0.1006  & 0.0664  & 0.0321  & 0.2724  & 0.1764  & 0.3397  & 0.2320  \\
          & MetaEmb & 0.1264  & 0.1105  & 0.1458  & 0.1501  & 0.2798  & 0.1663  & 0.0083  & 0.0045  & 0.2724  & 0.1764  & 0.3397  & 0.2320  \\
          & GAR   & 0.1372  & 0.1072  & 0.0177  & 0.0198  & 0.2487  & 0.1483  & 0.0284  & 0.0222  & 0.2490  & 0.1574  & 0.3398  & 0.2342  \\
          & ALDI  & 0.1405  & 0.1173  & 0.0207  & 0.0156  & 0.2647  & 0.1512  & 0.0359  & 0.0191  & 0.2724  & 0.1764  & 0.3397  & 0.2320  \\
\cmidrule{2-14}          & USIM  & 0.1346  & 0.1164  & 0.1458  & 0.1501  & 0.2654  & 0.1589  & 0.0164  & 0.0117  & 0.2724  & 0.1764  & 0.3397  & 0.2320  \\
          & MI-GNN & 0.1510  & 0.1322  & 0.1434  & 0.1409  & 0.0046  & 0.0022  & 0.0032  & 0.0028  & \underline{0.3078}  & \underline{0.2046}  & 0.3488  & 0.2407  \\
          & Wang et.al & 0.1887  & 0.1523  & 0.1304  & 0.1222  & 0.2552  & 0.1533  & 0.0513  & 0.0321  & 0.3011  & 0.2032  & 0.3165  & 0.1972  \\
          & ColdLLM & \underline{0.2046}  & \underline{0.1559}  & \underline{0.1594}  & \underline{0.1665}  & \underline{0.3353}  & \underline{0.2026}  & \underline{0.1035}  & \underline{0.0828}  & 0.2695  & 0.1726  & \underline{0.3795}  & \underline{0.2630}  \\
\cmidrule{2-14}          & \model & \textbf{0.2515} & \textbf{0.2124} & \textbf{0.1638} & \textbf{0.1765} & \textbf{0.3981} & \textbf{0.2578} & \textbf{0.1604} & \textbf{0.1221} & \textbf{0.3393} & \textbf{0.2299} & \textbf{0.3807} & \textbf{0.2718} \\
    \bottomrule
    \end{tabular}
  \label{tab:main}
\end{table*}%

In this section, we conduct extensive experiments on cold-start recommendation datasets and online A/B tests, aiming to answer the following questions.
\textbf{RQ1:} Does \model outperform contemporary state-of-the-art cold-start recommendation models in overall, warm, and cold recommendations?
\textbf{RQ2:} How does the \model adapt to LLMs with different parameters?
\textbf{RQ3:} How does \model achieve efficiency compared to other baselines?
\textbf{RQ4:} How does \model perform in real-world industrial recommendations?

\subsection{Experimental Setup}
\subsubsection{Datasets} We conduct offline experiments on two widely used datasets: \textbf{CiteULike}\footnote{\url{https://github.com/js05212/citeulike-a}}~\cite{wang2013citeulike}, containing 5,551 users, 16,980 articles, and 204,986 interactions, and \textbf{ML-10M}\footnote{\url{https://grouplens.org/datasets/movielens/10m}}~\cite{harper2015movielens}, comprising 6,9797 users, 1,0258 items, and 5,093,135 interactions. More details about datasets will be shown in Appendix~\ref{Dataset Details}.

\subsubsection{Compared Baselines}
To assess the effectiveness of our proposed \model, we compare \model with twelve leading-edge cold-start recommendation models, which can be categorized into three main groups.
(i) Dropout-based embedding simulation models: \textbf{DropoutNet}~\cite{volkovs2017dropoutnet}, \textbf{MTPR}~\cite{du2020mtpr}, \textbf{Heater}~\cite{zhu2020heater},and \textbf{CLCRec}~\cite{wei2021clcrec}. (ii) Generative-based embedding simulation models: \textbf{DeepMusic}~\cite{van2013deepmusic}, \textbf{MetaEmb}~\cite{pan2019metaemb}, 
\textbf{GAR}~\cite{chen2022gar}, and \textbf{ALDI}~\cite{huang2023aldi}.
(iii) Interaction simulation models: \textbf{USIM}~\cite{liu2024fine}, \textbf{MI-GNN}~\cite{wang2024mutual}, \textbf{Wang et.al}~\cite{wang2024large} and \textbf{ColdLLM}~\cite{huang2024large}. To further verify the universality of \model, we verify these models on three widely used recommendation backbones: \textbf{MF}~\cite{rendle2009bprmf}, \textbf{NGCF}~\cite{wang2019ngcf}, and \textbf{LightGCN}~\cite{he2020lightgcn}. More details about the implementation setting will be shown in Appendix~\ref{baseline_detail}.

\subsubsection{Implementation Setting}
For \model training, we used AdamW~\cite{loshchilov2017decoupled} with a learning rate of $1\times10^{-4}$ and a batch size of 8. The top-k value was set to 20. In enhancing cold embedding with the distribution process, the learning rate remained $1\times10^{-4}$. For a fair comparison, we used LLaMA2-7B~\cite{Llama} as the base LLM, consistent with ColdLLM.
More details about the implementation setting are shown in Appendix~\ref{Implementation Dtails}.

\subsubsection{Evaluation Metrics}
Our evaluation encompasses the overall, cold, and warm recommendation performance, adopting a widely adopted full-ranking evaluation approach~\cite{he2020lightgcn,huang2023aldi}. 
Specifically, we employ Recall@$K$ and NDCG@$K$ as our primary metrics, where $k=20$. Task description can be seen in Appendix~\ref{Task Descriptions}.

\subsection{Offline Evaluation (RQ1)}
The offline comparison of overall, warm, and cold recommendations between \model and twelve baselines is presented in Table ~\ref{tab:main}. From the results, we have the following observations.

\textbf{\model exhibits significant advantages over other baselines}. From the results, we can find that \model consistently outperforms existing approaches across three backbones, particularly delivering substantial improvements over LightGCN. 
Specifically, for overall recommendations, \model averagely outperforms the best baseline by 13.28\% and 2.13\% for Recall, and 20.08\% and 2.82\% for NDCG on CiteULike and ML-10M, respectively.
This enhancement underscores the effectiveness of \model based on the ``Text-to-Distribution'' paradigm.

\textbf{The LLM-based baseline ColdLLM achieves the best performance among other baselines.} Across different backbones, ColdLLM consistently stands out as the strongest baseline for both overall and cold-start recommendation tasks on the CiteULike and ML-10M datasets. This highlights the substantial impact of world knowledge embedded in large language models in enhancing the performance of cold-start recommendations.

\textbf{Dropout-based embedding simulators perform less effectively than the other two types of methods in overall and warm recommendation scenarios.} One possible reason is that dropout-based models overlook the distinction between warm behavior embeddings and cold content embeddings. Generative-based models address this discrepancy at the representation aspect, while interaction simulation methods directly eliminate this difference at the interaction aspect.

\subsection{Generalization Study (RQ2)}
\begin{figure}[t] 
    \centering
    \includegraphics[width=\columnwidth]{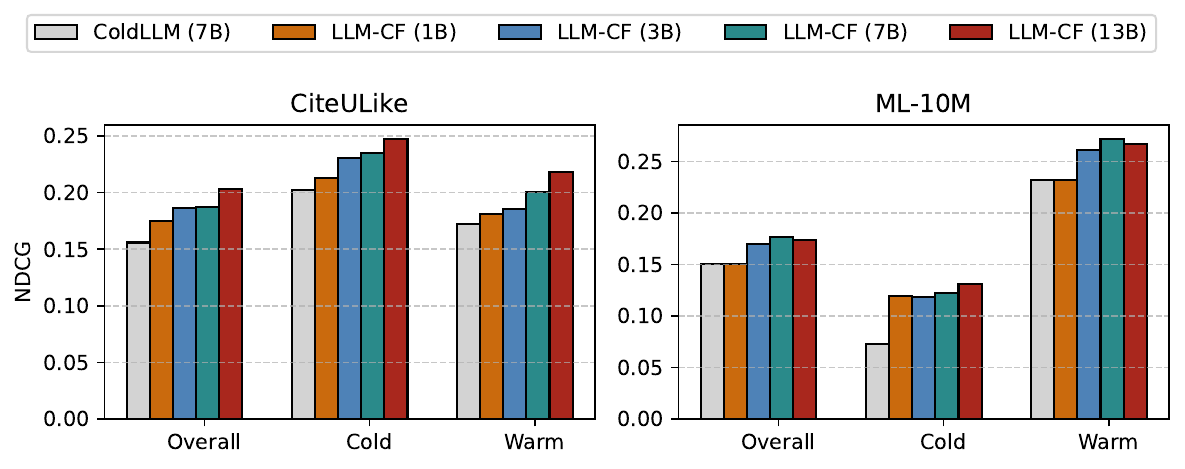} 
    \caption{Performance of \model across different LLMs compared to ColdLLM (LLaMA2-7B).}
    \label{fig:compartion}
\end{figure}

To evaluate \model's generalization ability across different LLM backbones, we conduct experiments on four base LLMs with varying parameter sizes: LLaMA3-1B, LLaMA3-3B, LLaMA2-7B, and LLaMA2-13B, and compare them against ColdLLM (LLaMA2-7B). The results are presented in Figure~\ref{fig:compartion}. From the results, we have the following observations.

\textbf{\model demonstrates strong performance across all LLM backbones even when using small-size LLMs.} Our experiments with small-size LLM (e.g., LLaMA3-1B) reveal that \model not only matches but also frequently surpasses the performance of the current state-of-the-art baseline. This remarkable outcome highlights the effectiveness and robustness of our FilterLLM framework based on the ``Text-to-Distribution'' paradigm. 

\textbf{As the parameter size of the LLM increases, the performance of \model tends to show improvements.} This may be attributed to the enhanced language understanding capabilities that come with the growing model scale. This trend also highlights \model's ability to effectively leverage the increased representational capacity of larger LLMs, translating it into superior recommendation performance. 

\subsection{Efficiency Study (RQ3)}
\label{Efficiency Study}

To evaluate the efficiency superiority of \model, we compared it with existing LLM-based models on model training time, average token length of LLM, and model inference time for \model and each baseline. More details about the experimental design are shown in Appendix~\ref{Experimental Design}. From the results presented in Table~\ref{tab:efficiency}, we can draw the following observations. 

\textbf{\model consistently demonstrates a significantly shorter model training time compared to ColdLLM.}
Specifically, we can find that \model's training time is nearly half of ColdLLM's, highlighting its superior training efficiency. This improvement is primarily attributed to the use of item-wise training data in \model, as opposed to the interaction-wise training data used by ColdLLM, which greatly reduces the computational requirements during training.

\textbf{\model also achieves remarkably faster inference times than other LLM-based models.}
Compared with ColdLLM, \model is 38.84 times faster on CiteULike and 36.53 times faster on ML-10M. Further, \model has 161.53 times faster on average than Wang et.al.
Two key factors drive this substantial improvement, the significant reduction in the average input token length and \model's ability to generate a user distribution for an item in a single forward pass, eliminating the need for multiple inferences. 
\begin{table}[t]
  \centering
  \caption{Model efficiency comparison results.}
  \renewcommand{\arraystretch}{1.15} 
  \resizebox{\linewidth}{!}{%
    \begin{tabular}{cc|ccc}
    \toprule
    \multicolumn{2}{c|}{Methods} & Training Time & Token Length & Inference Time \\
    \midrule
    \multirow{3}[2]{*}{\begin{sideways}CiteULike\end{sideways}} & Wang et.al~\cite{wang2024large}  &   Close-sourced    &   301.23    & 42.78 min (16.58$\times$)\\
          & ColdLLM~\cite{huang2024large} &   10.40 h    &  273.69     & 100.20 min (38.84$\times$)\\
    \cmidrule{2-5}      & \model &    \textbf{5.51 h}   &  \textbf{72.87}     & \textbf{2.58 min} (1$\times$) \\
    \midrule
    \multirow{3}[2]{*}{\begin{sideways}ML-10M\end{sideways}} & Wang et.al~\cite{wang2024large}  &    Close-sourced   &  593.50   & 530.21 min (306.47$\times$)\\
          & ColdLLM~\cite{huang2024large} &  17.33 h    &    541.77   & 63.21 min (36.53$\times$)\\
    \cmidrule{2-5}      & \model &   \textbf{5.23 h}    &  \textbf{102.49}     & \textbf{1.73 min} (1$\times$) \\
    \bottomrule
    \end{tabular}%
  }
  \label{tab:efficiency}%
\end{table}

\subsection{Online Evaluation (RQ4)}

\begin{table}[t]
  \centering
  \caption{Online evaluation results of \model.}
  \resizebox{\linewidth}{!}{ 
    \begin{tabular}{lcccc}
    \toprule
    A/B Test & \multicolumn{1}{c}{Cold-PV} & \multicolumn{1}{c}{Cold-PCTR} & \multicolumn{1}{c}{Cold-GMV} & \multicolumn{1}{c}{LLM Infer. Time} \\
    \midrule
    vs.ALDI &     +12.42  &  +7.17     &    +26.58\%   & - \\
    vs.ColdLLM &    +5.13\%   &    +3.93\%   &   +10.86\%    &  -97.12\%\\
    \bottomrule
    \end{tabular}%
  }
  \label{tab:A/B test}%
\end{table}%
To evaluate \model in an industrial setting, we conducted a two-month online A/B test on Alibaba's platform, which has 0.3 billion daily users and 1 million new items added per day. Users were randomly divided into two equal groups for the A/B test, and \model was compared against ALDI~\cite{huang2023aldi} and ColdLLM~\cite{huang2024large}. The evaluation metrics are detailed in Appendix \ref{Evaluation Metrics}. The results presented in Table~\ref{tab:A/B test} lead to the following key observations.

\textbf{\model demonstrates significant improvements in cold-start recommendation scenarios across multiple metrics,} outperforming the ALDI baseline with 12.42\%, 7.17\%, and 26.58\% gains in Cold-PV, Cold-PCTR, and Cold-GMV, respectively. Compared to ColdLLM, it shows relative improvements of 5.13\% in Cold-PV, 3.93\% in Cold-PCTR, and 10.86\% in Cold-GMV. These results demonstrate \model's effectiveness in boosting both user engagement and commercial outcomes.

\textbf{\model significantly enhances LLM inference efficiency compared to ColdLLM. }Specifically, \model requires only 2.88\% of ColdLLM's inference time. This substantial improvement is due to the transition from ``Text-to-Judgement'' to ``Text-to-Distribution'', which eliminates the need for repeated LLM inferences for each item and reduces the needed input token length.

\section{Conclusion}
LLM-based cold-start recommendation remains a critical challenge in billion-scale scenarios due to its reliance on the ``Text-to-Judgment'' paradigm. To address this, we propose a novel ``Text-to-Distribution'' paradigm, enabling LLMs to predict user interaction distributions for an item in a single inference. Building on this, we introduce FilterLLM, a framework that scales LLM inference for large-scale cold-start recommendations. Deployed on Alibaba’s platform for two months, FilterLLM serves as the backbone for cold-start recommendations. Extensive experiments validate its effectiveness and efficiency in both offline and online evaluations.


\bibliographystyle{ACM-Reference-Format}
\clearpage
\bibliography{sample-base}

\appendix
\section{Model Analysis Details}
\subsection{Pseudocode of \model}
\label{Pseudocode}
\begin{algorithm}[htbp]
    \renewcommand{\algorithmicrequire}{\textbf{Input:}}  
    \renewcommand{\algorithmicensure}{\textbf{Output:}} 
    \caption{The overall workflow of \model.}  
    \label{alg:framework}
    \begin{algorithmic}[1]
        \REQUIRE Item content $\gC$; User set $\gU$; Cold item $\gI_{c}$; Warm item $\gI_{w}$; Interaction $\gH$; Pretrained LLM weight $\theta_{LLM}$; Added user vocabulary embeddings $Z_{u}$.
        \ENSURE Cold item embeddings $\bm{E}^{(c)}_i$.
        
         \# Collaborative-Driven User Vocabulary Initialization

        \STATE $\bm{E}_{u},\bm{E}_{i}  \leftarrow \argmin_{\bm{E}_{u},\bm{E}_{i}} {\gL_{Token}}$
        \STATE $Z_u \leftarrow \bm{E}_{u}$;// Initialize the added token embedding with item oriented embeddings

        \# \model Training
        \WHILE{Until coverage}
        \STATE $\mathcal{L}_\text{Distrib} \leftarrow \{\gH,\gI_{w}\}$; //Calculate Distribution Learning Loss with Eq.(12)
        \STATE$\mathcal{L}_\text{Guid} \leftarrow \{\bm{E}_{i},\gI_{w}\}$; //Calculate Behavior
        Guiding Loss with Eq.(13)
        \STATE$\mathcal{L} \leftarrow \gL_\text{Distrib} + \lambda \cdot\mathcal{L}_\text{Guid} $
        \STATE $(\theta_{LLM}, Z_{u}) \leftarrow (\theta_{LLM}, Z_{u}) - \lambda\nabla\mathcal{L}$; //optimize the LLM and user vocabulary.
        \ENDWHILE

        \#Interaction Sampling
        \FOR{$i \in \gI_{c}$}
        \STATE Sample interactions $\hat{\gH}_i$ for cold item $i$ with Eq.(7),(8);
        \STATE $\hat{\gH} \leftarrow \hat{\gH} \cup \hat{\gH}_i$;
        \ENDFOR

        \#Cold Item Embeddings Optimization
        \STATE Optime the Cold Item Embeddings $\bm{E}^{(c)}_i$ with Eq.(5);
    \end{algorithmic}  
\end{algorithm}
\subsection{Complexity Analysis}
\label{Complexity Analysis}

 Let's start with the complexity of transformer structure~\cite{vaswani2017attention}. The transformer structure mainly contains two parts, self-attention and feed-forward neural network (FFN). Assuming the length of the input tokens is $L$, then the time complexity of self-attention and FFN are $O(L^{2}\cdot d)$ and $O(L\cdot d^{2})$, respectively. So the time complexity of the transformer structure is $O(n \cdot(L^2\cdot d+L\cdot d^2))$, where $n$ is the number of total layers. When ``text-to-judgment'' methods inferring, they will repeat $K$ times LLM forward process for one item, where $K$ is the size of the candidate set. So the complexity of ``text-to-judgment'' methods to generate interactions for a single item is $O (K \cdot n \cdot(L^2\cdot d+L\cdot d^2))$. As for our \model, it only needs one forward process and a top-k selection process, so the total time complexity of \model is $O(n \cdot(L^2\cdot d+L\cdot d^2) + |\gU|\log|\gU|)$. Considering that ``text-to-judgment'' methods and \model have different input lengths, the ratio of time complexity is $\frac{L_{2}^{2}+L_{2}d}{L_{1}^{2}+L_{1}d}K$, where $L_{2}$ and $L_{1}$ are average inputs length of ``text-to-judgment'' methods and \model.

\section{Experimental Details}
\subsection{Task Descriptions}
\label{Task Descriptions}
\paragraph{\textbf{Restrict Item Cold-start Recommendation.}}
This paper focuses on the most challenging strict cold-start problem, from the view of item cold-start, where the cold items lack any historical behaviors. Under this constraint, the warm items and the cold items lead to two different ways of recommendation. The warm items are recommended with historical user sequences, which are usually encoded into behavior embeddings. Formally, the warm recommendation can be defined as:
\begin{equation}
    \hat{y}^{(w)}_{ui} =
        R(\ve_{u}, \text{Emb}_{cf}(\vs_i), \vc_i), \ i \in \gI_w,
\end{equation}
where $\text{Emb}_{cf}(\cdot)$ denote the collaborative filtering function for behavior embedding. However, the user sequence set of the cold item is empty, making the cold items recommendation to be organized with the following formula:
\begin{equation}
    \hat{y}^{(c)}_{ui} =
        R(\ve_{u}, \text{Emb}_{cf}(\phi), \vc_i), \ i \in \gI_c.
\end{equation}
Thus the restricted cold-start recommendation problem turns to recommend the above warm items and cold items well.

\subsection{Dataset Details}
\label{Dataset Details}
We evaluate the \model's performance on cold-start items using the CiteULike and MovieLens datasets.

\begin{itemize}[leftmargin=*]    
    \item \textbf{CiteULike}\footnote{\url{https://github.com/Zziwei/Heater--Cold-Start-Recommendation/tree/master/data}}~\cite{wang2013citeulike} The dataset contains 5,551 users, 16,980 articles, and 204,986 interactions. On CiteULike, registered users create scientific article libraries and save articles for future reference. The articles are represented by 200-
    dimensional vectors as item content features for non-LLM-based methods. And followed \cite{huang2024large} LLM-based methods are only provided with the title of papers.
    
    \item \textbf{MovieLens}\footnote{\url{https://grouplens.org/datasets/movielens/1m}}~\cite{harper2015movielens} One of the most well-known benchmark data. We consider interactions where the user rating is greater than or equal to 4 as positive interactions and remove all negative interactions. After this filtering, the dataset consists of 69,797 users, 10,258 items, and 3,662,349 interactions. Titles and genres are used as item content in our experiment. The articles are represented by 200-dimensional vectors as item content features for non-LLM-based methods. 
\end{itemize}

For each dataset, following previous works~\cite{huang2023aldi}, 20\% items are designated as cold-start items, with interactions split into a cold validation set and testing set (1:1 ratio). Records of the remaining 80\% of items are divided into training, validation, and testing sets, using an 8:1:1 ratio.

\subsection{Baseline Details}
\label{baseline_detail}
To assess the effectiveness and universality of \model, we conducted a comparative analysis with 12 leading-edge models in the domain of cold-start recommendations. This comparison was carried out across two distinct datasets. The models we benchmarked against include three main groups: (i) Dropout-based embedding simulation models: \textbf{DropoutNet}~\cite{volkovs2017dropoutnet}, \textbf{MTPR}~\cite{du2020mtpr}, \textbf{Heater}~\cite{zhu2020heater}, and \textbf{CLCRec}~\cite{wei2021clcrec}. (ii) Generative-based embedding simulation models: \textbf{DeepMusic}~\cite{van2013deepmusic}, \textbf{MetaEmb}~\cite{pan2019metaemb}, 
\textbf{GAR}~\cite{chen2022gar}, and \textbf{ALDI}~\cite{huang2023aldi}.
(iii)Interaction simulation models: 
\textbf{USIM}~\cite{liu2024fine}, \textbf{MI-GNN}~\cite{wang2024mutual}, \textbf{Wang et.al}~\cite{wang2024large} and \textbf{ColdLLM}~\cite{huang2024large}. 

\begin{itemize}[leftmargin=*]
\item \textbf{DeepMusic} utilizes deep neural networks to model the mean squared error difference between generated and warm embeddings.
\item \textbf{MetaEmb} trains a meta-learning-based generator for rapid convergence.
\item \textbf{GAR} generates embeddings through a generative adversarial relationship with the warm recommendation model.
\item \textbf{ALDI} employs distillation, using warm items as "teachers" to transfer behavioral information to cold items, referred to as "students".
\item \textbf{DropoutNet} enhances cold-start robustness by randomly discarding embeddings.
\item \textbf{MTPR} generates counterfactual cold embeddings considering dropout and Bayesian Personalized Ranking (BPR).
\item \textbf{Heater} improves DropoutNet by using a mix-of-experts network and considering embedding similarity.
\item \textbf{CLCRec} models cold-start recommendation using contrastive learning from an information-theoretic perspective.

\item \textbf{USIM} proposes a user behavior imagination framework with a reinforcement learning strategy for cold items.

\item \textbf{MI-GNN} adopts pair-wise mutual information to generate informative interactions for cold items before the GNN convolution.

\item \textbf{Wang et.al} utilizes an LLM to analyze user interaction history and evaluate preferences for a pair of cold items, capturing semantic interactions.

\item \textbf{ColdLLM} adopts a filtering-and-refining approach, using a fine-tuned LLM to match users with target cold items and query potential interactions.
\end{itemize}

\label{Experimental Design}
\subsection{Implementation Dtails}
\label{Implementation Dtails}
We implement the baselines using their official implementation. Specifically, for GAR, we use
the updated version provided in the official repository, which is evaluated under the same CLCRec settings as used in our paper. The embedding dimension is set to 200 for non-LLM-based models. The experiment was conducted on four NVIDIA A100 Tensor Core GPUs with 80GB of memory. The max epoch of our \model is set to 20 and the \model weight corresponding to the lowest loss is selected for evaluation. Following ~\cite{huang2024large}, we use full updating to optimize the item embeddings in the main experiment.
\subsection{Experimental Design of Efficiency Study}
When calculating the model training time, the setup is consistent with the main experiment. For the model inference time, to clearly highlight the differences in inference efficiency across models, we exclude variations in memory consumption caused by differing input prompt lengths, which could lead to discrepancies in the maximum batch size the LLM can process. To ensure consistency, for models using open-source LLMs (ColdLLM, \model), we perform inference on a single A100 GPU with a batch size of 1 and a maximum input length of 1024. For models using closed-source LLMs (Wang et al.), we query GPT-3.5 one by one.

\section{More Experimental Results}
\subsection{Ablation Study}
\begin{figure}[t] 
    \centering
    \includegraphics[width=\columnwidth]{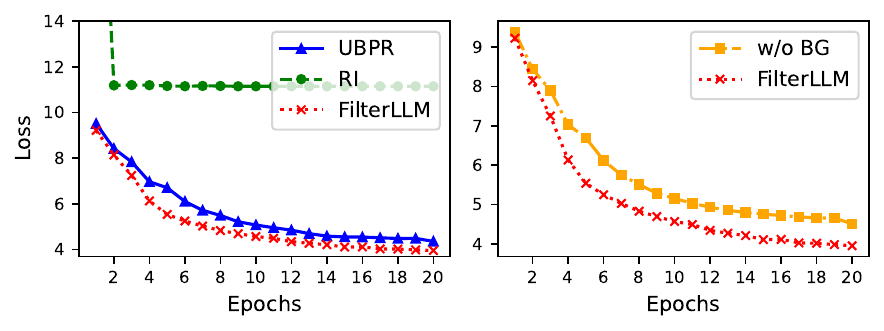} 
    \caption{Loss curve of \model and its variants on ML-10M.}
    \label{fig:Ablation Study}
\end{figure}

To validate the effectiveness of the individual components in our \model, we compared the full model against three variants:(i)\textbf{\textit{w/o BG}} removes the behavior guiding loss. (ii) \textbf{\textit{RI}} randomly initializes the user vocabulary embeddings. (iii) \textbf{\textit{UBPR}} unitizes the original BPR loss to initialize the user vocabulary embeddings. The results are illustrated in Table ~\ref{tab:ablation}. Furthermore, to examine and evaluate the training efficiency, we plotted the loss curve depicting the reduction of loss over epochs of \model and three variants in Figure~\ref{fig:Ablation Study}.
\begin{table}[t]
  \centering
  \caption{Ablation study results between \model with its three variants on ML-10M.}
    \begin{tabular}{l|rr|rr|rr}
    \toprule
    \multicolumn{1}{c|}{\multirow{2}[2]{*}{Variant}} & \multicolumn{2}{c|}{Overall Rec.} & \multicolumn{2}{c|}{Cold Rec.} & \multicolumn{2}{c}{Warm Rec.} \\
          & \multicolumn{1}{c}{Recall} & \multicolumn{1}{c|}{NDCG} & \multicolumn{1}{c}{Recall} & \multicolumn{1}{c|}{NDCG} & \multicolumn{1}{c}{Recall} & \multicolumn{1}{c}{NDCG} \\
    \midrule
    w/o BG & 0.1521 & 0.1595 & 0.1412 & 0.1058 & 0.3563 & 0.2431 \\
    RI    & 0.1423 & 0.1514 & 0.0223 & 0.0164 & 0.3377 & 0.2341 \\
    UBPR  & 0.1559 & 0.1629 & 0.1407 & 0.1032 & 0.3587 & 0.2447 \\
    \midrule
    \model  & \textbf{0.1638} & \textbf{0.1765} & \textbf{0.1604} & \textbf{0.1221} & \textbf{0.3807} & \textbf{0.2718} \\
    \bottomrule
    \end{tabular}%
  \label{tab:ablation}%
\end{table}%

\textbf{Effectiveness of Collaborative-Driven User Initialization.}
The decline in performance for \textbf{\textit{RI}} demonstrates the importance of the right way to initialize the user vocabulary embeddings. Moreover, while \textbf{\textit{UBPR}} outperforms \textbf{\textit{RI}}, it still falls short compared to \model. This highlights the importance of designing tailored initialization methods specifically for items. Additionally, as observed from the loss reduction curves, both \textbf{\textit{RI}} and \textbf{\textit{UBPR}} exhibit slower convergence compared to \model, further underscoring the significance of our proposed Collaborative-Driven User Vocabulary Initialization in facilitating effective model training.

\textbf{Effectiveness of Behavior Guiding.} The performance drop observed in \textbf{\textit{w/o BG}} indicates that the item embeddings learned from collaborative filtering effectively compensate for the LLM's limitations in capturing high-order user-item relationships, thereby improving the model's performance. Moreover, the loss curve of \textbf{\textit{w/o BG}} declines more slowly compared to \model, and its final loss remains higher than that of \model. This demonstrates that the behavior guiding loss facilitates more efficient learning for the model.

\subsection{Parameter Study Results}
\begin{figure}[h] 
    \centering
    \includegraphics[width=\columnwidth]{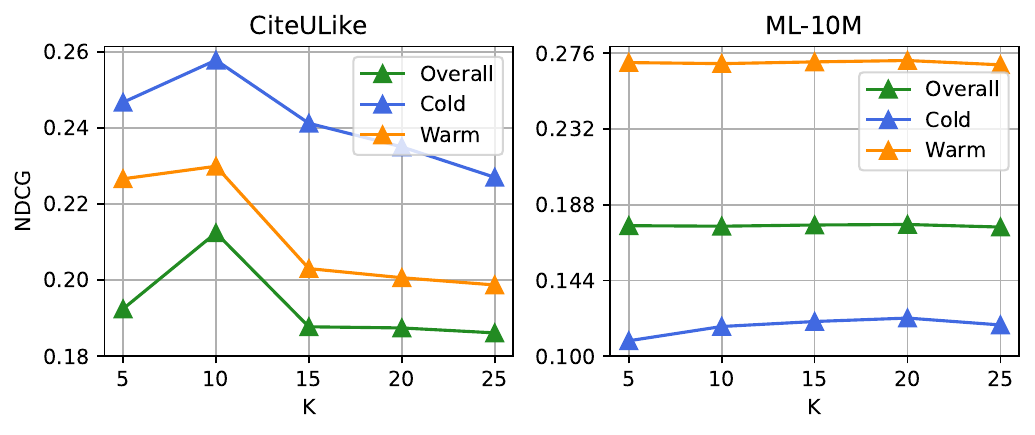} 
    \caption{Parameter study of \model on $K$.}
    \label{fig:K}
\end{figure}

\begin{figure}[h] 
    \centering
    \includegraphics[width=\columnwidth]{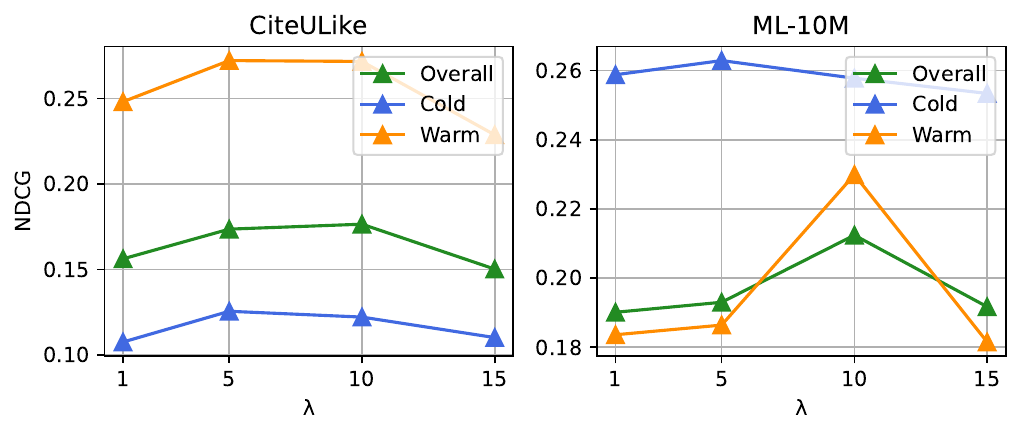} 
    \caption{Parameter study of \model on $\lambda$.}
    \label{fig:lambda}
\end{figure}

In this subsection, we conducted a parameter analysis on top-K number $K$ and behavior guiding weight $\lambda$ on dataset CiteULike and ML-10M. The results are shown in Figure \ref{fig:K} and Figure \ref{fig:lambda}. From these results, we can draw the following conclusions.

\textbf{$K$ has a significant impact on CiteULike datasets.} In the CiteULike dataset, when K=10, the performance for warm, cold, and overall interactions is optimal, with a large difference compared to other values. This is likely due to the small number of users and items in the dataset, where more simulated interactions affect performance.

\textbf{$K$ primarily affects cold interactions on ML-10M datasets.} In the ML-10M dataset, we observe that as K increases, the changes in performance for warm and overall interactions are minimal compared to those for cold interactions. This could be because of the large volume of data in ML-10M, where although the noise increases with larger K, the amount of noise for warm items is still relatively small compared to the actual interactions, so the performance for both overall and warm items remains almost unchanged.

\textbf{Moderate $\lambda$ values are required for both the CiteULike and ML-10M datasets.} As observed in the results, both overall and cold-start performance peak at $\lambda=10$ and $\lambda=5$, respectively, in both datasets. This may be because a smaller $\lambda$ prevents the model from effectively learning from item behavior embeddings, while an excessively large $\lambda$ may cause the model to overfit to noise in these embeddings.

\section{Industrial Evaluation Metrics.}
\label{Evaluation Metrics}
We employed three tailored metrics to assess the performance of \model against existing baselines:
\begin{itemize}[leftmargin=*]
\item  Page Views (Cold-PV): The number of user clicks during the cold period. It can be formulated as:
\[
\text{Cold-PV} = \sum_{i \in \mathcal{I}} \sum_{u \in \mathcal{U}} \mathbf{1}_{\text{click}(u, i)},
\]
where \(\mathcal{I}\) is the set of items, \(\mathcal{U}\) is the set of users, and \(\mathbf{1}_{\text{click}(u, i)}\) is an indicator function that equals 1 if user \(u\) clicks item \(i\), and 0 otherwise.

\item  Page Click-Through Rate (Cold-PCTR): The ratio of clicks to impressions during the cold period. It can be expressed as:
\[
\text{Cold-PCTR} = \frac{\sum_{i \in \mathcal{I}} \sum_{u \in \mathcal{U}} \mathbf{1}_{\text{click}(u, i)}}{\sum_{i \in \mathcal{I}} \sum_{u \in \mathcal{U}} \mathbf{1}_{\text{impression}(u, i)}},
\]
where \(\mathbf{1}_{\text{impression}(u, i)}\) is an indicator function that equals 1 if item \(i\) is shown to user \(u\), and 0 otherwise.

\item Gross Merchandise Value (Cold-GMV): The total value of user purchases during the cold period. It is given by:
\[
\text{Cold-GMV} = \sum_{i \in \mathcal{I}} \sum_{u \in \mathcal{U}} \text{purchase}(u, i) \cdot \text{price}(i),
\]
where \(\text{purchase}(u, i)\) equals 1 if user \(u\) purchases item \(i\), and 0 otherwise, and \(\text{price}(i)\) is the price of item \(i\).
\end{itemize}


\end{document}